\documentclass[twocolumn,showpacs,amsmath,amssymb,aps,prl,floatfix,nofootinbib,superscriptaddress]{revtex4}

\usepackage{graphicx,graphics,times}% Include figure files

\usepackage{bm}% bold math

\begin{document}

\title{Why does the Standard GARCH(1,1) model work well?}

\author{G. R. Jafari}
\email{gjafari@gmail.com} \affiliation{Department of Physics, Shahid
Beheshti University, Evin, Tehran 19839, Iran}
\affiliation{Department of Nano-Science, IPM, P. O. Box 19395-5531,
Tehran, Iran}

\author{A. Bahraminasab}
\email{abahrami@ictp.it} \affiliation{Department of Physics,
Lancaster University, Lancaster LA1 4YB, United Kingdom}
\affiliation{The Abdus Salam International Center for Theoretical
Physics (ICTP) Strada Costiera 11, I-34100 Trieste, Italy}

\author{P. Norouzzadeh}
\email{payam.norouzzadeh@ua.ac.be} \affiliation{Department of
Physics, University of Antwerp, Groenenborgerlaan 171, B-2020
Antwerpen, Belgium}

\begin{abstract}
The AutoRegressive Conditional Heteroskedasticity (ARCH) and its
generalized version (GARCH) family of models have grown to encompass
a wide range of specifications, each of them is designed to enhance
the ability of the model to capture the characteristics of
stochastic data, such as financial time series. The existing
literature provides little guidance on how to select optimal
parameters, which are critical in efficiency of the model, among the
infinite range of available parameters. We introduce a new criterion
to find suitable parameters in GARCH models by using Markov length,
which is the minimum time interval over which the data can be
considered as constituting a Markov process. This criterion is
applied to various time series and results support the known idea
that GARCH(1,1) model works well.
\end{abstract}

%\date{\today}
\maketitle
%%%%%%%%%%%%%%%%%%%%%%%%%%%%%%%%%%%%%%%%%%%%%%%%%%%%%%%%%%%%%%%%%%%%%%%%%%%%%%%%%%%%%
\section{Introduction}
The ARCH model \cite{Engle1} and standard GARCH model
\cite{Bollerslev1} are now not only widely used in the Foreign
Exchange (FX) literature \cite{Bollerslev2} but also as the basic
framework for empirical studies of the market micro-structure such
as the impact of news \cite{Goodhart1} and government interventions
\cite{Goodhart2, Peiers}, or inter- and intra-market relationships
\cite{Engle2, Baillie}. Due to basic similarities between the
mentioned models of different constants, here we focus our
discussion on financial context.

The main assumption behind this class of models is the relative
homogeneity of the price discovery process among market participants
at the origin of the volatility process. Volatility is an essential
ingredient for many applied issues in finance. It is becoming more
important to have a good measure and forecast of short-term
volatility, mainly at the one to ten day horizon. Currently, the
main approaches to compute volatilities are by historical indicators
computed from daily squared or absolute valued returns, by
econometric models such as GARCH. In other words, the conditional
density of one GARCH process can adequately capture the information.
In particular, GARCH parameters for the weekly frequency
theoretically derived from daily empirical estimates are usually
within the confidence interval of weekly empirical estimates
\cite{Drost}. It is interesting to note that despite the extensive
list of models which now belong to the ARCH family \cite{Andersen,
Bera, Bollerslev3}, the existing literature provides little guidance
on how to select optimal $q$ and $p$ values in a GARCH ($q$,$p$)
model. The two parameters $q$ and $p$ in GARCH model refer to
history of the return and volatility of time series, respectively.
However, there are some criteria to find the suitable $q$ and $p$.
Pagan and Sabu \cite{Pagan}, suggest a misspecified volatility
equation that can result in inconsistent maximum likelihood
estimates of the conditional mean parameters \cite{Solibakke}.
Further West and Cho \cite{West} show how appropriate GARCH model
selection can be used to enhance the accuracy of exchange rate
volatility forecasts. However, there is an almost ubiquitous
reliance on the standard GARCH (1,1)-type model in the applied
literature. In this paper, we suggest a new approach to find optimal
values $q$ and $p$. In our approach, a fundamental time scale is the
Markov time scale, $t_M$, which is the minimum time interval over
which the data can be considered as constituting a Markov process.
As our starting point, we proceed by stressing on the role of $q$
and $p$ in the GARCH model and their similarity to the Markov length
scale in stochastic processes. There are reported examples of
measuring Markov length for various data \cite{Friedrich1, Siefert}.
Recently Tabar et al., \cite{Bhattacharyya}, have used the dynamics
of the Markov length for predicting earthquakes. Moreover, Renner et
al. \cite{Renner} have shown that Markov time scale in returns of
the high frequency (minutely) US Dollar - German Mark exchange rates
in the one-year period October '92 to September '93 is larger than 4
minutes.

Generally, we check whether the data follow a Markov chain and, if
so, measure the Markov length scale, $t_M$. Such a given process
with a degree of randomness or stochasticity may have a finite or an
infinite Markov length scale. Specifically, the Markov length scale
is the minimum length interval over which the data can be considered
as a Markov process. The main goal of this paper is to utilize the
mentioned similarity between the role of $q$ and $p$ in the GARCH
model and Markov time scale in stochastic processes, and introduce a
novel method to estimate optimal GARCH model parameters of daily
data.
%@@@@@@@@@@@@@@@@@@@@@@@@@@@@@@@@@@@@@@@@@@@@@@@@@@@@@@@@@@@@@@@@@@@@@

\section{GARCH Based Volatility Models}

The GARCH model, which stand for Generalized AutoRegressive
Conditional Heteroscedasticity, is designed to provide a volatility
measure like a standard deviation that can be used in financial
decisions concerning risk analysis, portfolio selection and
derivative pricing. The GARCH models have become important tools in
the analysis of time series data, particularly in financial
applications. This model is especially useful when the goal of the
study is analyze and forecast volatility.

Let the dependent variable be labeled by $r_t$, which could be the
return on an asset or portfolio. The mean value $\mu$ and the
variance $\sigma^2$ will be defined relative to past information
set. Then, the return $r$ in the present will be equal to the mean
value of $r$ (that is ,the expected value of $r$ based on past
information) plus the standard deviation of $r$ times error term for
the present period.

The challenge is to specify how the information is used to forecast
the mean and variance of the return, conditional on the past
information. The primary descriptive tool was the rolling standard
deviation. This is the standard deviation calculated using a fixed
number of the most recent observations. It assumes that the variance
of tomorrow's return is an equally weighted average of the squared
residual from the past days. The assumption of equal weights seems
unattractive as one would think that the more recent events would be
more relevant and therefore should have higher weights. Furthermore
the assumption of zero weights for observations more than one period
old, is also unattractive. The GARCH model for variance looks like
this \cite{Bollerslev1}:
%%%%%%%%%%%%%
\begin{eqnarray}
r_{t}&=&\mu+\sigma_{t}\varepsilon_{t}\cr\nonumber\\
\sigma^2_t&=&\alpha_0 +\alpha_1 r_{t-1}^2+ ....... + \alpha_p r_{t-p}^2\cr\nonumber\\
 &\hspace{2mm}&+\beta_1 \sigma_{t-1}^2+ ....... +\beta_q \sigma_{t-q}^2
\end{eqnarray}
%%%%%%%%%%%%%%%%%%
where $r_{t}$ denotes the returns and $\varepsilon_{t}\sim iid
\hspace{1mm} N(0,1)$. The constants $\alpha_0$ up to $\alpha_p$ and
$\beta_1$ up to $\beta_q$ must be estimated. Upating simply requires
knowing the previous forecast $\sigma ^{2}$ and residual. In the
GARCH(1,1) model, the weights are respectively ($1-\alpha_1-\beta_1,
\alpha_1, \beta_1$) and the long run average variance is $\surd
\frac{\alpha_0}{1-\alpha_1-\beta_1}$. It should be noted that this
only works if $\alpha_1+\beta_1<1$, and only really makes sense if
the weights are positive requiring $\alpha_1\geq0$, $\beta_1\geq0$
and $\alpha_0>0$. The described GARCH model is typically called the
GARCH(1, 1) model. The first parameter in GARCH(q, p) models refers
to the number of autoregressive lags or ARCH terms appearing in the
equation, while the second parameter refers to number of moving
average lags, which here is often called the number of GARCH terms.
Sometimes models with more than one lag are needed to find good
variance forecasts. For estimating parameters of an equation like
the GARCH(1, 1) when the only variable on which there are data is
$r_{t}$ we can use maximum likelihood approach by substituting
conditional variance for unconditional variance in the normal
likelihood and then maximize it with respect to the parameters. The
likelihood function provides a systematic way to adjust the
parameters $\alpha_0, \alpha_1$ and $\beta_1$ to give the best fit.

%==============================================regen

\section{Estimation of the GARCH parameters by calculating Markov length scale}

We begin by describing the procedure that lead to find optimal GARCH
parameters $(p,q)$ based on the (stochastic) data set. As the first
step we check whether the return (volatility) of data follows a
Markov chain and, if so, measure the Markov length scale $t_M$ and
find $p$ (also $q$) equal to $t_M$. As is well-known, a given
process with a degree of randomness or stochasticity may have a
finite or an infinite Markov length scale. The Markov length is the
minimum time interval over which the data can be considered as a
Markov process. To determine the Markov length $t_M$, we note that a
complete characterization of the statistical properties of
stochastic fluctuations of a quantity $r$ in terms of a parameter
$t$ requires the evaluation of the joint probability distribution
function (PDF) $P_n(r_1, t_1;  \dots; r_n, t_n)$ for an arbitrary
$n$, the number of the data points. If the phenomenon is a Markov
process, an important simplification can be made, as the n-point
joint PDF, $P_n$, is generated by the product of the conditional
probabilities $p(r_{i+1}, t_{i+1}|r_i, t_i)$, for $i = 1, \dots , n
- 1$. A necessary condition for a stochastic phenomenon to be a
Markov process is that the Chapman- Kolmogorov (CK) equation
\cite{Risken},
\begin{eqnarray}\label{ck}
  P(r_2,t_2|r_1,t_1)=  \int \hbox{d} r_3
  P(r_2,t_2|r_3,t')P(r_3, t'| r_1,t_1).
\end{eqnarray}
should hold for any value of $t'$ in the interval $t_2 < t' < t_1$.
One should check the validity of the CK equation for different $r_1$
by comparing the directly-evaluated conditional probability
distributions $p(r_2, t_2|r_1, t_1)$ with the ones calculated
according to right side of equation (1). The simplest way to
determine $t_M$ for stationary or homogeneous data is numerical
calculation of following quantity,
\begin{eqnarray}\label{s}
S = |p(r_2, t_2|r_1, t_1)-\int dr_3p(r_2, t_2|r_3, t') p(r_3,
t'|r_1, t_1)|, % \nonumber\\
\end{eqnarray}
for given $r_1$ and $r_2$, in terms of a time interval, for example,
$t' -
 t_1$. After that, $t_M = t' -t_1$ if the $S$ value vanishes (regarding the
possible errors in estimating $S$). To find $q$ we repeat the above
procedure for volatility series.

A simple formula like as $a_n=a_{n-1}+a_{n-2}$ refers that to obtain
the present value we need to consider two successive past values. In
fact, the Markov length of such processes is equal to $2$. In this
way, one can interpret $q$ and $p$ in GARCH ($q$,$p$) model as
Markov time scales for returns and volatilities time series,
respectively. Therefore, we suggest a new framework to find the $q$
and $p$ values in GARCH models through calculating Markov time scale
for both return and volatility series. Based on new approach the
Markov lengths in return and volatility series show us how many
steps in these series we need to go back to get a good description
of the process. Additionally, we can interpret Markov length as a
kind of memory of the data. Knowing this memory helps us to
understand how present values of return and volatility are affected
by past valuse.
\begin{figure}[t] % fig 1
\includegraphics[width=8.1cm,height=7.1cm,angle=0]{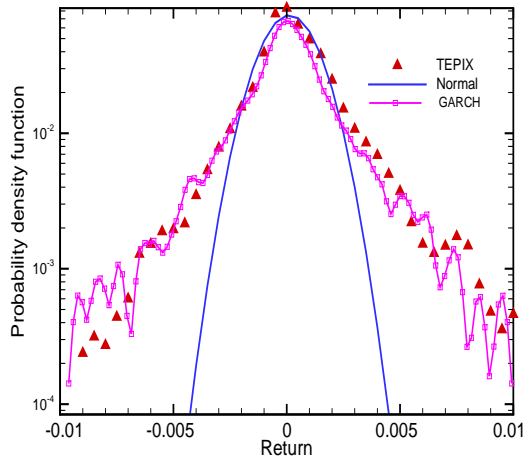}
\caption{Comparison of the empirical probability distribution of
TEPIX returns with the fitted normal distribution and GARCH(1,1)
model.}
\end{figure}

%-------------------------------------------------------------

\begin{figure}[t] % fig 2
\includegraphics[width=9.1cm,height=13.1cm,angle=0]{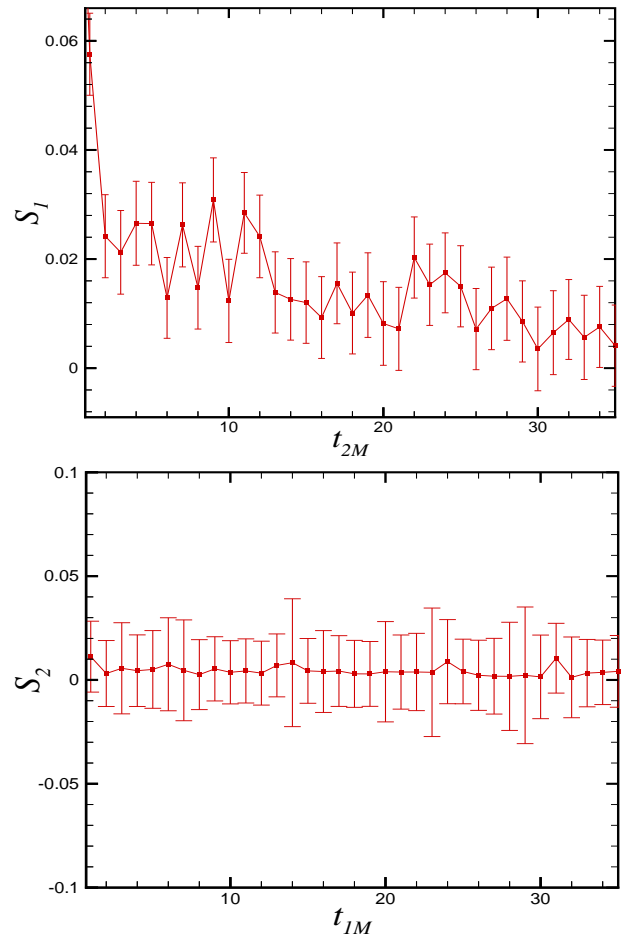}
\caption{The $S$ values of daily prices of Brent Oil along with
their statistical errors for return (upper graph) and volatilities
(lower graph) in period 20 October 1985 to 24 September 2006.}
\end{figure}

Using this approach, we have considered some time series related to
financial markets and commodities. We have calculated their Markov
time scales for both returns and volatilities. Results are presented
in Table 1. Taking into account results, indicates that Markov time
scale of returns and volatilities for most of such data is equal 1.
This implies that GARCH (1,1) model can be a well established model
for financial modeling and descriptions. The calculated likelihood
values for various GARCH ($q$,$p$) models support our approach.
There is a good agreement between likelihood results and what Markov
time scale approach suggests. Normally, the GARCH (1,1) model works
better than other GARCH models and now it seems that the standard
model has obtained another confirmation and could well capture the
most important features of these kind of data in general. An
interesting feature of our approach is that it works for both normal
and fat-tail distributions. We have presented skewness, kurtosis and
Hurst exponent values of selected time series in Table 1 to show
statistical properties and fat-tailness of them \cite{jafari}. They
show different degrees of fat-tailness. Based on presented values in
Table 1 the selected time series are far from normal distribution
and have fat-tail distributions features and are very close to
normal distributions. For normal distributions Hurst exponent is
equal to 0.5 and greater exponents indicate fat-tailness. S\&P500
and dow-jones indices together brent oil prices have lower degrees
of fat-tailness compared to TEPIX and NZX. For a better comparison
of the return distribution of TEPIX as an example of fat-tail
distributions with a normal distribution and GARCH(1,1) model, the
empirical PDF is presented in Fig.1 along with the fitted normal and
GARCH(1,1) PDFs.

%-------------------------------------------------------------

 We have used above formalism for some daily time series such as $S\&P500$ (20
October 1982 to 24 November 2004), Dow-Jones( 4 January 1915 to 20
February  1990), New Zealand Exchange (NZX-7 January 1980 to 30
December 1999), Brent oil prices (4 January 1982 to 24 September
2006) and Tehran Price Index (TEPIX-20 May 1994 to 18 March 2004).

The value of $S$ in Eq. (3) has been calculated for two time series:
returns and volatilities. In fig. 1 the results of $S$ values
related to daily prices of Brent Oil  along with their statistical
errors for different time scales are shown. The interesting point is
that our calculations show that Markov time scale for the daily
return and volatility series in Brent oil prices is $q=1$ and $p=2$,
respectively.
%%%%%%%%%%%%%%%%%%%%%%%%%%%%%%%%%%%%%%%%%%%%%%%%%%%%%%%%%%%%%%%%%%%%%%%%%%%

%%%%%%%%%%%%%%%%%%%%%%%%%%%%%%%%%%%%%%%%%%%%%%%%%%%%%%%%%%%%%%%%%%%%%%%%%%%

\begin{table}
\caption{\label{Tb1}The estimated values GARCH process parameters
using Markov length scale approach}
\medskip
\begin{tabular}{|c|c|c|c|c|c|}
  \hline\hline
  \hspace{1mm}Time series \hspace{1mm}& Skewness & Kurtosis &Hurst exponent&\hspace{2mm}$q$\hspace{2mm} & \hspace{2mm}$p$ \hspace{2mm}\\\hline
  S$\&$P 500  &1.93&45.10&0.44 &1 & 1 \\\hline
  Dow-Jones&-0.25&17.43&0.53 &1&  1  \\\hline
  NZX &-1.43&27.29&0.61&1& 1 \\\hline
  TEPIX &0.72&18.50& 0.74&1& 1  \\\hline
  Brent Oil &-1.25&34.99&0.51&2 &1   \\\hline\hline
\end{tabular}
\end{table}

In summary, we introduced a new criterion to find optimal $q$ and
$p$ in GARCH family models. We have shown that a fundamental time
scale in our approach is the Markov length, $t_M$, which is the
minimum time interval over which the data can be considered as
constituting a Markov process. This criterion support the success of
standard GARCH (1,1) model to describe the majority of time series.
In fact we suggest calculating Markov length for both return and
volatility series to find proper parameters of GARCH models before
proceeding to describe such processes and now we have a new method
to select optimal $q$ and $p$ parameters in a GARCH ($q$, $p$)
models.

\section{acknowledgment}

We would like to thank A. T. Rezakhani for his useful comments and
discussions.

 %%%%%%%%%%%%%%%%%%%%%%%%%%%%%%%%%%%%%%%%%%%%%%%%%


\begin{thebibliography}{99}

\bibitem{Engle1} Engle R. F., (1982), Autoregressive conditional heteroskedasticity
with estimates of the variance of U. K. in ation, Econometrica, 50,
987.

\bibitem{Bollerslev1} Bollerslev, T. (1986). Generalized autoregressive conditional
heteroskedasticity. Journal of Econometrics 31, 307.

\bibitem{Bollerslev2} Bollerslev T, Engle R F and Nelson D B, ARCH models Handbook of
Econometrics, vol 4, ed R F Engle and D M Mcfadden (1994).

\bibitem{Goodhart1} Goodhart C. A. E. and Figliuoli L., (1991), Every minute counts in
financial markets, Journal of International Money and Finance, 10,
23; Goodhart C. A. E., Hall S. G., Henry S. G. B., and Pesaran B.,
(1993), Journal of Applied Econometrics, 8, 113.

\bibitem{Goodhart2} Goodhart C. A. E. and Hesse T., (1993), Central Bank Forex
intervention assessed in continuous time, Journal of International
Money and Finance, 12(4), 368.

\bibitem{Peiers} Peiers B., (1994), A high-frequency study on the relationship
between central bank intervention and price leadership in the
foreign exchange market, unpublished manuscript, Department of
Economics, Arizona State University.

\bibitem{Engle2} Engle R. F., Ito T., and Lin W. L., (1990), Meteor showers or heat
waves? Heteroskedastic intra-daily volatility in the foreign
exchange market, Econometrica, 58, 525.

\bibitem{Baillie} Baillie R. T. and Bollerslev T., (1990), Intra day and inter market
volatility in foreign exchange rates, Review of Economic Studies,
58, 565.

\bibitem{Drost} Drost F. and  Nijman T., (1993), Temporal aggregation of
garch processes, Econometrica, 61, 909..

\bibitem{Andersen} Andersen T. G. and Bollerslev T., (1994), Intraday seasonality and
volatility persistence in foreign exhcnage and equity markets,
Kellogg Graduate School of Management, Northwestern University,
working paper 186, 1 30.

\bibitem{Bera} Bera A K and Higgins M L, ARCH models: properties, estimation and
testing, J. Econ. Surv. 4 305 (1993).

\bibitem{Bollerslev3}Bollerslev, T. and Domowitz, I. (1993). Trading patterns and prices
in the interbank foreign exchange market. Journal of Finance 48,
1421.

\bibitem{Pagan} Pagan A.R. and Sabu H.,(1992), Consistency Tests for Hetroskedastic
and Risk Models, Estudios Economics 7,30.

\bibitem{Solibakke} Solibakke P.B., Applied Financial Economics, Volume 11, Number
5,(2001), 539.

\bibitem{West} West K. D. and Cho D., (1994), The predictive ability of several
models of exchange rate volatility, NBER technical working paper,
152, 1.

\bibitem{Friedrich1} Friedrich R. and  Peinke J., Phys. Rev. Lett. {\bf 78}, 863
(1997); Friedrich R., et al., Phys. Lett. A 271, 217 (2000);
Friedrich R., Peinke J., and Renner C., Phys. Rev. Lett. 84, 5224
(2000); Kriso S., et al., Phys. Lett. A 299, 287 (2002); Siegert S.,
Friedrich R. and Peinke J., Phys. Lett. A 243, 275 (1998); Jafari G.
R., Fazeli S.M., Ghasemi F., Vaez Allaei S.M.,Reza Rahimi Tabar M.,
Irajizad A.,Kavei G., Phys. Rev. Lett. {\bf 91}, 226101 (2003);
Friedrich R., Zeller J., and  Peinke J., Europhysics Letters 41, 153
(1998); Ghasemi F., Peinke J., Sahimi M. and Reza Rahimi Tabar M.,
Eur. Phys. J. B  47, 411 (2005).

\bibitem{Siefert} Siefert M., Kittel A., Friedrich R. and Peinke J., Euro. Phys. Lett.
61, 466 (2003);

\bibitem{Bhattacharyya} Bhattacharyya P. and Chakrabarti B. K. , {\it{Modelling Critical and
Catastrophic Phenomena in Geoscience: A Statistical Physics
Approach}}, (Springer, Berlin, 2006), p. 281;

\bibitem{Renner} Renner C, Peinke J, Friedrich R, Physica A 2001, 298, 211.

\bibitem{Risken} Risken H., {\it{The Fokker Planck Equation}} (Springer, Berlin,
1984)

\bibitem{jafari}  Jafari G R, Movahed M S, Fazeli S M, Rahimi Tabar M Reza and
Masoudi S F, J. Stat. Mech. (2006) P06008.

\end{thebibliography}
\end{document}